\documentclass[twocolumn,english,prl,showpacs,superscriptaddress]{revtex4}

\usepackage{amssymb}

\makeatletter

\usepackage{graphicx}

\usepackage{babel}

\makeatother

\begin{document}

\title{Orthogonally-Driven Superconducting Qubit in Circuit QED}

\author{M.J.~Storcz\footnote{These authors contributed equally to this work.}}

\affiliation{Physics Department, CeNS and ASC,
Ludwig-Maximilians-Universit\"at, Theresienstrasse 37, 80333
Munich, Germany}

\author{M.~Mariantoni$ ^{*}$}

\affiliation{Walther-Meissner-Institut, Bayerische Akademie der
Wissenschaften, Walther-Meissner-Strasse 8, 85748 Garching,
Germany}

\author{H.~Christ}

\affiliation{Max-Planck Institute for Quantum Optics,
Hans-Kopfermann-Strasse 1, 85748 Garching, Germany}

\author{A.~Emmert}

\affiliation{Walther-Meissner-Institut, Bayerische Akademie der
Wissenschaften, Walther-Meissner-Strasse 8, 85748 Garching,
Germany}

\author{A.~Marx}

\affiliation{Walther-Meissner-Institut, Bayerische Akademie der
Wissenschaften, Walther-Meissner-Strasse 8, 85748 Garching,
Germany}

\author{W.D.~Oliver}

\affiliation{MIT Lincoln Laboratory, 244 Wood Street, Lexington,
Massachussets 02420, USA}

\author{R.~Gross}

\affiliation{Walther-Meissner-Institut, Bayerische Akademie der
Wissenschaften, Walther-Meissner-Strasse 8, 85748 Garching, Germany}

\author{F.K.~Wilhelm}

\affiliation{Physics Department, CeNS and ASC,
Ludwig-Maximilians-Universit\"at, Theresienstrasse 37, 80333
Munich, Germany}

\author{E.~Solano}

\affiliation{Physics Department, CeNS and ASC,
Ludwig-Maximilians-Universit\"at, Theresienstrasse 37, 80333
Munich, Germany}

\affiliation{Secci\'{o}n F\'{\i}sica, Departamento de Ciencias,
Pontificia Universidad Cat\'{o}lica del Per\'{u}, Apartado 1761,
Lima, Peru}

\date{\today}

\begin{abstract}
We consider a superconducting charge qubit coupled to distinct
orthogonal electromagnetic field modes belonging to a coplanar
waveguide resonator and a microstrip transmission line.
This architecture allows the simultaneous implementation of a
Jaynes-Cummings and anti-Jaynes-Cummings dynamics, a resonant
method for generating mesoscopic qubit-field superpositions and
for field-state reconstruction. Furthermore, we utilize this setup
to propose a field measurement technique that is, in principle,
robust due to a fast pre-measurement to qubit dephasing and field relaxation.
\end{abstract}


\maketitle

In recent years, superconducting quantum circuits have
demonstrated key elements required for quantum information
processing~\cite{Nielsen:2000:a}, including the possibility to
prepare a desired qubit state, coherently manipulate it, read it
out, and perform preliminary conditional gate
operations~\cite{Makhlin:2001:a,Devoret:2004:a,Yamamoto:2003:a}.
In analogy to quantum-optical cavity QED, superconducting charge
and flux qubits have been coupled to on-chip microwave
resonators~\cite{Wallraff:2004:a,Johansson:2006:a},
and universal two-qubit gates mediated by a single cavity mode
have been proposed~\cite{Blais:2004:a}. In
such an architecture, the scaling of the system would require a
homogeneous coupling of many qubits with the same cavity mode and
a means to address each logical qubit individually.
Prototypical advanced manipulation schemes have already been
implemented in quantum-optical systems. In trapped ion
experiments, tuned lasers can be switched
between single-ion carrier excitations and ion-motion JC
dynamics~\cite{Leibfried:2003:a}. Also, in 3D microwave cavities,
a flying atom can perform local rotations in Ramsey zones before
and after entering the cavity, in which a Jaynes-Cummings~(JC)
interaction takes place~\cite{Raimond:2001:a}. In the emerging field of circuit
QED~\cite{Blais:2004:a,Wallraff:2004:a}, a higher level of
addressability and control is also desirable~\cite{GJohansson:2006:a}. For instance, it is
important to enable controlled \emph{intracavity} qubit rotations,
while keeping a switchable coupling to the cavity modes, aiming at
\mbox{\emph{intercavity}} qubit-qubit transfer of information.
This implies the necessity to generate nonclassical field states
and the possibility of measuring them via rapid qubit operations.
These requirements could be simultaneously met if each qubit were coupled to two (or more)
independent modes with orthogonal field~polarizations.

\begin{figure}[t]
\begin{center}
\includegraphics[width=0.6\columnwidth,clip=]{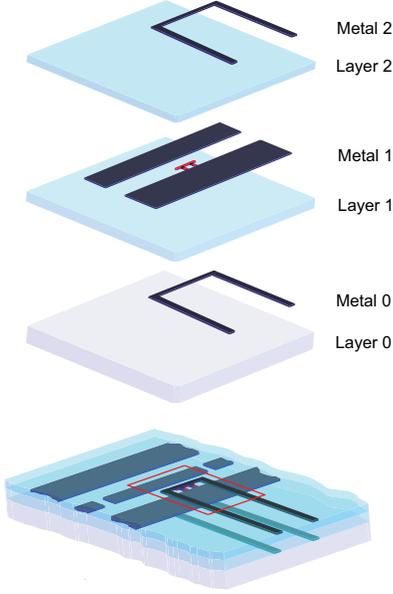}
\end{center}
\caption{(Color online)~Sketch of the proposed setup, including the blow-up of different layers, where  the CPB is coupled to orthogonal fields produced by a CWG and a MTL.} \label{Figure1}
\end{figure}

In this Letter, we propose an architecture consisting of a
superconducting charge qubit coupled to the quantized,
discrete-mode spectrum of a quasi-1D coplanar~waveguide~(CWG)
resonator, here called {\it cavity}~\cite{Wallraff:2004:a},
and to a multi-layer microstrip transmission~line~(MTL), which
will be utilized to tune the qubit-cavity resonance.
This archetypical construction ideally subjects the qubit to two
orthogonal electromagnetic fields and constitutes the system for
our theoretical investigations (See Fig.~\ref{Figure1}). We will show that strongly driving
the MTL with coherent field pulses modulates the strength of the
qubit-cavity interaction and, in the strong-driving
limit~\cite{Solano:2003:a}, mediates the emergence of a
simultaneous JC and anti-JC dynamics in the system, allowing the
generation of nonclassical qubit-cavity
states~\cite{Brune:1996:a}. Furthermore, we propose the
measurement of relevant observables of the microwave cavity field
via a protocol that utilizes the MTL as an independent tool for
read-out of the qubit population~\cite{Lougovski:2005:a}. Finally,
we will demonstrate that this technique is, in principle, robust
due to a fast qubit-cavity pre-measurement to the presence of qubit dephasing and field decoherence~processes.

The Hamiltonian describing the interaction between a single
Cooper-pair box~(CPB) charge qubit, the second
harmonic of the undriven CWG resonator, and the MTL (see Fig.~\ref{Figure1}), driven with a
propagating coherent state acting as an AC gate charge, can be
written as~\cite{H:RM:extra}
\begin{eqnarray}
 \hat{\bar{H}} & = & - \ 2 E^{}_{\rm C} \left( 1 - 2
  n^{\rm DC}_{\rm C} \right) \hat{\bar{\sigma}}^{}_z - \frac{E^{}_{\rm J}
\left( \Phi^{}_{\rm x} \right)}{2} \hat{\bar{\sigma}}^{}_x + \hbar
\omega^{}_{\rm C} \hat{a}^{\dag}_{2} \hat{a}^{}_{2} {}
\nonumber\\
& & + \ \hbar g^{}_{\rm QC} \hat{\bar{\sigma}}^{}_z \left(
\hat{a}^{\dag}_{2} + \hat{a}^{}_{2} \right) + 4 E^{}_{\rm C}
n^{\rm AC}_{\rm M} ( t ) \hat{\bar{\sigma}}^{}_z .
\label{Hamiltonian:RM}
\end{eqnarray}
Here, $E^{}_{\rm C} = e^2 / C^{}_{\Sigma}$~represents the charging
energy of the CPB~($C^{}_{\Sigma}$ is the total box capacitance),
$n^{\rm DC}_{\rm C}$ is the number of pairs induced by the DC gate
voltage through the CWG cavity, $E^{}_{\rm J}$ is the qubit
Josephson energy, which can be tuned by an external quasi-static
flux bias $\Phi^{}_{\rm x}$, applied through a adequately
engineered loop, $\{
\hat{a}^{\dag}_{2} , \hat{a}^{}_{2} \}$ are the bosonic creation
and destruction operators relative to the second harmonic of the
CWG resonator, $\omega^{}_{\rm C}$ and $g^{}_{\rm QC}$ are the
corresponding angular frequency and qubit-cavity~vacuum Rabi
coupling, respectively, $n^{\rm AC}_{\rm M} ( t )$ is the number
of pairs induced by the AC gate voltage applied via the MTL, and
$\{ \hat{\bar{\sigma}}^{}_x , \hat{\bar{\sigma}}^{}_z \}$ are
Pauli spin-$1/2$ operators. With $4 E^{}_{\rm C} n^{\rm AC}_{\rm
M} ( t ) = ( C_{\rm QM} V_{\rm QM}^0 / 2e ) |\beta|
 \cos ( \omega_{\rm M} t + \vartheta^{}_{\beta} ) $, we can
define $\hbar g^{\beta}_{\rm QM} \equiv ( C_{\rm QM} V_{\rm QM}^0
/ 2e ) |\beta|$, where $g_{\rm QM}$ is the qubit-MTL coupling
strength, $C^{}_{\rm QM}$ and $V_{\rm QM}^0$ are the associated
capacitance and vacuum gate voltage, $\beta = | \beta | \exp i
\vartheta^{}_{\beta}$ is the amplitude of the driven coherent
field state $| \beta \rangle$, $\omega^{}_{\rm M}$~its angular
frequency, and $\vartheta^{}_{\beta}$ a phase. The direct coupling
between the CWG resonator and the MTL is suppressed due to the
high-degree of isolation between their respective modes.
The MTL magnetic field distribution~\cite{Itoh:1989:a}
is predominantly tangential to the CPB loop, suppressing unwanted
jitter of the $\hat{\bar{\sigma}}^{}_{\rm x}$ term
in~Eq.~(\ref{Hamiltonian:RM}) due to MTL noise.

Working in the eigenbasis~$\{ | {\rm g} \rangle , | {\rm e}
\rangle \}$ of the first two terms of Eq.~(\ref{Hamiltonian:RM}),
the system Hamiltonian takes the~form
\begin{eqnarray}
\hat{H} = && \!\!\! \frac{\Omega}{2} \hat{\sigma}^{}_z + \hbar
\omega^{}_{\rm C} \hat{a}^{\dag}_{2} \hat{a}^{}_{2} + \left[ \hbar
g^{}_{\rm QC} \left( \hat{a}^{\dag}_{2} + \hat{a}^{}_{2} \right) +
4 E^{}_{\rm C} n^{\rm AC}_{\rm M} ( t ) \right]  \nonumber \\
&& \!\!\! \times \left( \cos \theta \hat{\sigma}^{}_z - \sin
\theta \hat{\sigma}^{}_x \right) \, ,
\label{Hamiltonian:RM:Rotated}
\end{eqnarray}
where~$\{ \hat{\sigma}^{}_x , \hat{\sigma}^{}_z \}$ are Pauli
matrices in the~$\{ | {\rm g} \rangle , | {\rm e} \rangle
\}$~eigenbasis, \mbox{$\Omega = \sqrt{ E^{2}_{\rm J} + \left[ 4
E^{}_{\rm C} \left( 1 - 2 n^{\rm DC}_{\rm C} \right) \right]^2}$}
is the qubit level separation, and \mbox{$\theta = \arctan \left[
E^{}_{\rm J} / 4 E^{}_{\rm C} \left( 1 - 2 n^{\rm DC}_{\rm C}
\right) \right]$} is the mixing angle. Operating the qubit at the
degeneracy point, i.e., for $n^{\rm DC}_{\rm C} = 1 / 2$, under
complete resonance conditions~[$E^{}_{\rm J} / ( 2 \hbar ) =
\omega^{}_{\rm C} = \omega^{}_{\rm M}$], and
within a standard rotating-wave approximation (RWA) in the
interaction picture, Eq.~(\ref{Hamiltonian:RM:Rotated}) can be
rewritten as
\begin{eqnarray}
\hat{H}^{\rm I} = && \!\!\!\!\! \hbar g^{}_{\rm QC} \left(
\hat{\sigma}^{}_{+} \hat{a}^{}_{2} + \hat{\sigma}^{}_{-}
\hat{a}^{\dagger}_{2} \right) + \hbar g^{\beta}_{\rm QM}
\hat{\sigma}_x \, . \label{Hamiltonian:I:RM}
\end{eqnarray}

A more advanced application of Eq.~(\ref{Hamiltonian:I:RM}) is
obtained in the strong-driving limit, $g^{\beta}_{\rm QM} \gg
g^{}_{\rm QC}$. In this case, following
Ref.~\cite{Solano:2003:a}, we
utilize an additional interaction representation with respect to
the second term on the r.h.s. of Eq.~(\ref{Hamiltonian:I:RM}).
Decomposing $\sigma_{\rm\pm}=(\sigma_x\pm i\sigma_y)/2$
allows one to omit the quickly-precessing $\sigma_y$ term through
a second RWA, yielding the effective Hamiltonian
\begin{eqnarray}
\hat{H}_{\rm eff}& = & \hbar \frac{g^{}_{\rm QC}}{2} \left(
\hat{\sigma}^{}_{+} + \hat{\sigma}^{}_{-} \right) \left(
\hat{a}^{\dagger}_{2} + \hat{a}^{}_{2} \right) \, .
\label{Hamiltonian:R:eff}
\end{eqnarray}
This strong-driving limit results in a circuit-QED realization of
a simultaneously resonant JC and anti-JC dynamics.
It is characterized here by a large coupling $g^{}_{\rm QC} / 2$,
and it enables the generation of mesoscopic superposition states
between the qubit and the cavity field.
For example, taking the initial qubit-cavity state
to be $| {\rm g} , 0^{}_{\rm C} \rangle = \left( | + \rangle + | -
\rangle \right) | 0^{}_{\rm C} \rangle / \sqrt{2}$, where $| \pm \rangle$ are the
$\hat{\sigma}_x$ eigenstates of the qubit with eigenvalues $\pm
1$, the evolution associated with Eq.~(\ref{Hamiltonian:R:eff})
after an interaction time $t^{}_{\rm int }$ yields the following
Schr\"odinger cat state
\begin{eqnarray}
| \Psi_{\rm cat} \rangle = \frac{ ( | + \rangle | \alpha \rangle +
| - \rangle | - \alpha \rangle )}{\sqrt 2} , \label{cat:state}
\end{eqnarray}
with $\alpha = -i g^{}_{\rm QC} t^{}_{\rm int} / 2$.
Using realistic experimental parameters, e.g., $g^{}_{\rm QC}
\approx 100$~MHz~\cite{Schuster:2006:a} and
cavity decay rate $\kappa^{}_{\rm C} \approx 0.1$~MHz, would allow
the generation of a Schr\"odinger cat state with amplitude~$|
\alpha | = \sqrt{\langle \hat{n}^{}_{\rm cat} \rangle} = \left(
g^{}_{\rm QC} / 2 \kappa^{}_{\rm C} \right)^{1 / 3} \approx 8$ ($64$ photons),
obtained for a maximum interaction time $t^{}_{\rm int} = 1 /
\kappa^{\rm eff}_{\rm C}$, where \mbox{$\kappa^{\rm eff}_{\rm C} =
| \alpha |^{2}_{} \kappa^{}_{\rm C}$}. This amplitude compares well with the best values obtained in 3D
microwave cavity QED with circular Rydberg
atoms~\cite{Auffeves:2003:a} and previous proposals in the resonant~\cite{Liu:2005:a}  and dispersive regimes~\cite{Melo:2006:a}.

In order to generate successfully the specific state $| \Psi_{\rm cat}
\rangle$ inside the cavity, it is necessary to decouple the
\mbox{qubit-cavity} system after the desired interaction time $t^{}_{\rm int}$.
This may be implemented by applying a second strong coherent state to the MTL, with
decoupling frequency $\omega^{}_{\rm dec}$ and amplitude $| \gamma^{}_{\rm dec} |$,
that AC-Stark shifts the CPB by an amount~$\delta$. At this point,
a single-shot measurement of the qubit
 state $| {\rm g} \rangle$ ($| {\rm e}
\rangle$) would leave the field state in an even (odd) coherent
state $| \Psi^{\rm e}_{\alpha} \rangle = ( | \alpha \rangle + | -
\alpha \rangle) / \sqrt{1 + \exp \left( - | \alpha |^2 \right)}$
($| \Psi^{\rm o}_{\alpha} \rangle = ( | \alpha \rangle - | -
\alpha \rangle) / \sqrt{1 - \exp \left( - | \alpha |^2
\right)}$)~\cite{Solano:2003:a}. This qubit measurement may be
made, for instance, by driving a probing field via the
\emph{third} harmonic of the CWG resonator, which is strongly
detuned from the qubit transition frequency by an amount
$\omega^{}_{\rm C} / 2 \gg g^{}_{\rm QC}$, for a time~$t^{}_{\rm
meas}$, and thereby allowing an independent QND \mbox{read-out} of
the qubit population via the resultant phase shift of the probing
field~\cite{Wallraff:2005:a}.

The architecture proposed here could be applied directly to
implement conventional cavity field measurements, consisting of an initial
qubit-cavity pre-measurement in which the qubit acts as the
quantum probe with which the cavity field is entangled, followed
by a measurement of the qubit~\cite{Zurek:2003:a}.
Typically, one requires an interaction time sufficiently long
(order of the inverse interaction frequency) to entangle strongly
the cavity field with the qubit. Accordingly, one typically
desires to operate in the qubit-cavity strong-coupling regime to
minimize this interaction time and, thereby, reduce the noisy
action of decoherence~\cite{Leibfried:2003:a,Raimond:2001:a}.
In contrast to this common approach, we will show below an
alternative means to implement a measurement of the cavity field
with a relatively fast qubit-cavity pre-measurement (small
fraction of the inverse interaction time) and minimal action of
decoherence~\cite{Lougovski:2005:a}.

We consider now the even coherent state $| \Psi^{\rm e}_{\alpha} \rangle$ obtained after measuring
the qubit in the ground state $|g\rangle$, as seen from Eq.~(\ref{cat:state}). Through the MTL,
a \mbox{$\pi / 2$-pulse} of duration $t^{}_{\rm rot}$ and resonant
with the shifted qubit transition can be applied,
yielding $\hat{\rho}^{}_{\rm S} ( t^{}_{\rm S} ) = | +^{}_{\phi}
\rangle \langle +^{}_{\phi} | \otimes \hat{\rho}^{\rm
e}_{\alpha}$, with $| +^{}_{\phi} \rangle = ( | {\rm g} \rangle +
e^{i \phi} | {\rm e} \rangle ) / \sqrt{2}$. Here, $\phi$ is a
relative phase, $\hat{\rho}^{\rm
e}_{\alpha} = | \Psi^{\rm e}_{\alpha} \rangle \langle \Psi^{\rm
e}_{\alpha} |$, and $t^{}_{\rm S} =
t^{}_{\rm int} + t^{}_{\rm m} + t^{}_{\rm rot}$ may be set to
$t^{}_{\rm S} = 0$. At this point, if the aforementioned
decoupling pulse (frequency $\omega^{}_{\rm dec}$ and amplitude $|
\gamma^{}_{\rm dec} |$) applied via the MTL is turned off for a
time $t^{}_{\rm off}$, the qubit-cavity state~$\hat{\rho}^{}_{\rm
S} ( 0 )$ will evolve via a resonant JC dynamics for a
dimensionless time $\tau = g^{}_{\rm QC} t^{}_{\rm off}$.

In the presence of a dispersive bath, producing qubit dephasing,
and a thermal bath at zero temperature, inducing field
dissipation, the system dynamical equation becomes
\begin{displaymath}
 \frac{d}{d \tau} \hat{\rho}^{}_{\rm S} = \frac{1}{i \hbar}
 \left\lbrack \hat{H}^{}_{\rm QC} , \hat{\rho}^{}_{\rm S}
 \right\rbrack - \sum_i \frac{\gamma_i}{2} \left( \left\{
 \hat{A}_i^\dagger \hat{A}_i , \hat{\rho}^{}_{\rm S} \right\} - 2
 \hat{A}^{}_i \hat{\rho}^{}_{\rm S} \hat{A}_i^\dagger \right) \, ,
\end{displaymath}
where $\hat{H}^{}_{\rm QC} = \hbar g^{}_{\rm QC} \left(
\hat{\sigma}^{}_{+} \hat{a}^{}_{2} + \hat{\sigma}^{}_{-}
\hat{a}^{\dagger}_{2} \right)$, the braces denote
anti-commutators, and the sum runs over two indices corresponding
to qubit dephasing ($\hat{A}^{}_1 = \hat{\sigma}^{}_z ,
\gamma^{}_1 = \gamma^{}_{\phi}$) and cavity decay ($\hat{A}^{}_2 =
\hat{a}^{}_{2} , \gamma^{}_2 = \kappa^{}_{\rm C}$), respectively.
Above, we have neglected the energy relaxation of the qubit which
is assumed to happen on a longer timescale~\cite{Wallraff:2005:a}.
Calculating now $d P_{\rm e} ( \tau ) / d \tau = d \langle | {\rm
e} \rangle \langle {\rm e} | \rangle / d \tau = {\rm Tr}
\left\lbrack \dot{\hat{\rho}^{}}_{\rm S} | {\rm e} \rangle \langle
{\rm e} | \right\rbrack$, and after some algebra, we obtain
\begin{eqnarray}
\frac{d}{d \tau} P_{\rm e} ( \tau ) = \frac{1}{i \hbar}
\left\langle \left\lbrack | {\rm e} \rangle \langle {\rm e} | ,
\hat{H}^{}_{\rm QC} \right\rbrack \right\rangle \, .
\label{Ehrenfest}
\end{eqnarray}
This expression does not contain terms involving qubit dephasing
or field decay rates, which, ideally, are
eliminated by the expectation value calculated at time~$\tau$.
Evaluating~Eq.~(\ref{Ehrenfest}) in the limit
$\tau \rightarrow 0^+$ yields the useful result
\begin{eqnarray}
 \left\langle \hat{Y}^{}_{\phi} \right\rangle = \frac{d}{d \tau}
 P_{\rm e}^{+_{\phi}} ( \tau ) \Bigg|_{\tau \rightarrow 0^+} ,
 \label{Quadrature}
\end{eqnarray}
where $\hat{Y}^{}_{\phi} = \left( \hat{a}^{}_{2} e^{- i \phi} -
\hat{a}^{\dagger}_{2} e^{i \phi} \right) / 2 i$ is a field
quadrature, conjugate to the quadrature~$\hat{X}^{}_{\phi} =
\left( \hat{a}^{}_{2} e^{- i \phi} + \hat{a}^{\dagger}_{2} e^{i
\phi} \right) / 2$ obtained by replacing $\phi \rightarrow ( \phi
- \pi / 2)$. Equation~(\ref{Quadrature}) shows mathematically that the first
derivative of the measured excited-state qubit population
obtained at infinitesimally small interaction time contains
information about cavity-field observables with no influence of decoherence
processes. If the qubit is now prepared in the excited state $| \rm e
\rangle$, it is also possible to determine the mean photon number
of the cavity field, via
\begin{eqnarray}
 \label{photon_number}
 \left\langle \hat{n} \right\rangle = \left\langle \hat{a}^{\dag}
 \hat{a} \right\rangle = - \frac{1}{2} \frac{d^{2}}{d^{2} \tau}
 P_{\rm e}^{\rm e} ( \tau ) \Bigg|_{\tau \rightarrow 0^+} - 1.
\end{eqnarray}
Fast pre-measurements, plus the near unit visibility read-out of qubit populations
reported in Ref.~\cite{Wallraff:2005:a}, turns the proposed
measurement scheme into an alternative means to measure cavity-field observables with reduced noise disturbance.

Although Eqs.~(\ref{Quadrature}) and~(\ref{photon_number}) are
exact mathematical expressions, they do not represent a realistic
theoretical description of a measurement. In this sense, it is useful to consider an estimator for the derivatives in
Eqs.~(\ref{Quadrature}) and~(\ref{photon_number}) over short, but
non-zero, measurement times $\Delta \tau$ through the Taylor expansion: $P ( \tau + \Delta \tau ) = P ( \tau) + P' ( \tau ) \Delta \tau + P'' ( \tau ) ( \Delta \tau )^2 / 2 !
+ {\cal O} \lbrack (\Delta \tau)^3 \rbrack$. Subsequently, it can be shown that
\begin{eqnarray}
\label{Taylor} \frac{\Delta P_{\rm e}^{+_{\phi}} ( \tau )}{\Delta
\tau} \Bigg|_{\tau = 0} \!\!\! \approx \left\langle \hat{Y}^{}_{\phi}
\right\rangle \! - \!\! \bigg\lbrack 1+ \frac{ ( \kappa^{}_{\rm C} + \gamma_{\phi} ) } {4
g^{}_{\rm QC}} \, \left\langle \hat{Y}^{}_{\phi} \right\rangle  \bigg\rbrack
\Delta \tau \, ,
\end{eqnarray}
where the second term on the r.h.s.~is the dominant higher-order
contribution to Eq.~(\ref{Quadrature}), which, even for ideal
ensemble averaging, contains the field decay rate $\kappa^{}_{\rm
C}$ and the qubit dephasing rate $\gamma_{\phi}$.
Equation~(\ref{Taylor}) shows that in the strong-coupling regime,
$\{ \kappa_{\rm C} , \gamma_{\phi} \} < g_{\rm QC}$, or even in
the weak-coupling regime, $\{ \kappa_{\rm C} , \gamma_{\phi} \} >
g_{\rm QC}$, it is possible to identify a time
$\Delta \tau$ that is long enough to allow the readout of
$\langle \hat{Y}^{}_{\phi} \rangle$ and short enough to suppress
the effects of decoherence during the pre-measurement. This result
illustrates that, in principle, strong-coupling between the probe
and the system is not required to implement the proposed
measurement scheme.

Establishing an accurate
estimator with the prescribed accuracy for the ensemble averages
in Eqs.~(\ref{Quadrature}) and~(\ref{photon_number}) will, in
general, require many repetitions of the prescribed measurement.
There is a trade-off between the length of the physically
implemented $\Delta\tau$, the number of measurement repetitions,
and the strength of the qubit-cavity coupling required to achieve
a desired degree of noise immunity. In consequence, the proposed measurement technique consists of two clear steps. First, a fast pre-measurement allows the pointer (qubit) to encode the information about the system (cavity field), minimizing decoherence processes. Second, the readout of the pointer, happening typically over longer times, completes the quantum measurement procedure.

Finally, another useful application of our scheme is the
possibility to perform {\it full state reconstruction} of an unknown
field state $\hat{\rho} = | \Psi \rangle \langle \Psi |$ or,
equivalently, of its corresponding characteristic function $\chi (
\tilde{\alpha} ) = {\rm Tr} \left[ \hat{\rho} D ( \tilde{\alpha} )
\right]$. Here, $D ( \tilde{\alpha} )$ is the displacement
operator and $\tilde{\alpha}$ the complex amplitude in phase space
of an arbitrary coherent state~\cite{Cahill:1969:a}. For the sake
of convenience, a pure field state will be considered here, even though
these results remain valid for any arbitrary mixed state.
Following a similar protocol to the one outlined above, the qubit
can be initially prepared in the state~$| +^{\theta} \rangle = ( |
+ \rangle + e^{i \theta} | - \rangle ) / \sqrt{2}$, while the CWG
cavity, populated with the unknown field $| \Psi_{\rm C} \rangle$,
stays unperturbed. At this point, the interaction described in
Eq.~(\ref{Hamiltonian:R:eff}) can be turned on and the initial
state $| +^{\theta} \rangle |\Psi_{\rm C} \rangle$, after an
interaction time $t_{\rm d}$, evolves to
\begin{eqnarray}
| \Psi ( \tilde{\alpha} , \theta ) \rangle = \frac{ ( | + \rangle
D ( \tilde{\alpha} ) | \Psi_{\rm C} \rangle + e^{i \theta} | -
\rangle D ( - \tilde{\alpha} ) | \Psi_{\rm C} \rangle ) }{\sqrt 2}
\, , \label{characteristic:state}
\end{eqnarray}
with $\tilde{\alpha} =- i g_{\rm QC} t_{\rm d} / 2$. By measuring
the ground state qubit population $P_{\rm g} ( \tilde{\alpha} ,
\theta)$, given that the initial state was $| +^{\theta} \rangle$,
we can retrieve the characteristic function through
\begin{eqnarray}
\chi ( \alpha ) = \left\lbrack P_{\rm g} \left( \frac{\alpha}{2} ,
0 \right) - \frac{1}{2} \right\rbrack + i \left\lbrack P_{\rm g}
\left( \frac{\alpha}{2} , \frac{\pi}{2} \right) - \frac{1}{2}
\right\rbrack \, .
\end{eqnarray}
From this measured function $\chi ( \alpha )$, it is possible to
derive $\hat{\rho}$, or its associated Wigner
function~\cite{Melo:2006:a,Zou:2004:a}, via a Fourier
transform~\cite{Cahill:1969:a}, achieving a full-state
reconstruction.

One means of coupling a CPB to two quasi-orthogonal electric
fields is to make use of a multi-layer technology. In Fig.~\ref{Figure1}, dielectric
\emph{Layer 0} is the substrate for the entire structure, and the
MTL ground plane is \emph{Metal 0}. Dielectric \emph{Layer 1}
serves as a substrate for the CWG resonator-CPB structures, which
are made from \emph{Metal 1}. Dielectric \emph{Layer 2} supports
the MTL made from \emph{Metal 2}. The proper engineering of this structure
results in quasi-perpendicular electric fields at the CPB. These
fields are described by dyadic Green's functions and can be
evaluated numerically with the method of
moments~\cite{Collin:1990:a}. The results of our simulations
indicate about~$- 40$~dB isolation at $5$~GHz. For example, driving the MTL
with a coherent state of $\langle \hat{n} \rangle \approx
10^{3}$~photons would populate the CWG resonator with $\sim
0.1$~photons.

In conclusion, we have considered a CPB coupled to the orthogonal
modes of a CWG and MTL resonators. This architecture
allows the engineering of simultaneous JC and anti-JC dynamics, which we
utilized to generate mesoscopic entangled states of the CPB and the CWG.
It may be applied to the measurement of the field quadratures
using a measurement scheme requiring a relatively short
interaction time between the system (cavity) and the probe
(qubit). We have also shown that full state reconstruction of
unknown fields can be obtained. The multimode concepts described here should bring greater
flexibility to the field of circuit QED, in particular those
aiming at the systematic scaling-up of quantum gates and quantum
information devices.

This work was supported by DFG through SFB~631. The work at
Lincoln Laboratory was sponsored by the US DoD under Air Force
Contract No. FA8721-05-C-0002. ES acknowledges support of EU EuroSQIP project.


\begin{thebibliography}{50}
\bibitem{H:RM:extra}
Constant energy offsets, the cavity vacuum term, and negligible
second order terms have been~omitted.
\end{thebibliography}
\end{document}